\documentstyle{article}
\begin{document}
\rightline{IFUG-95-R-4}
\rightline{August 22, 1995}
\rightline{quant-ph/9509017}

\begin{center}{\Large{\bf Penning Trap and Vacuum Noise}\\

H. Rosu
\footnote{e-mail: rosu@ifug.ugto.mx}\\[2mm]
Instituto de F\'{\i}sica, Universidad de Guanajuato,\\
Apdo Postal E-143, 37150 Le\'on, Gto, M\'exico\\  }
\end{center}


\vskip 1cm

\begin{center}
{\bf Abstract}

A number of comments are provided on Rogers's model experiment to measure the
circular Unruh vacuum noise by means of a hyperbolic Penning trap inside a
microwave cavity. It is suggested that cylindrical Penning traps, being
geometrically simpler, and controlled almost at the same level of accuracy as
the hyperbolic trap, might be a better choice for such an experiment. Besides,
the microwave modes of the trap itself, of known analytical structure, can be
directly used in trying to obtain measurable results for such a tiny noise
effect.

\end{center}

\vskip 1cm


\hskip 2.7cm theories


\vskip 2cm


The physics of electromagnetic traps is an extremely vigorous discipline,
both theoretically and experimentally speaking \cite{p}.
Electrons in the Penning trap (PT), also known as electron geoniums and
associated mainly with $g-2$ experiments,
are simple physical systems offering
many advantages for measuring fundamental quantum field effects.
They have an internal dynamic behaviour which is
understood in great detail and they can be prepared and fully controlled
by means of modern laser spectroscopic methods. The PTs are characterized by
very high detection efficiency and high accuracy and tend to
become the favorite devices for studying experimentally the
theoretical results on radiative interactions, as well as fundaments of
quantum mechanics \cite{mt}.

On the other hand, vacuum noise is a concept vehiculated by people in
quantum field
 theory \cite{t}. If a quantum particle is following a classical
trajectory endowed with an acceleration parameter then because of the
{\em quantum vacuum} (or {\em quantum aether}) there could appear a radiative
 ``thermal" noise directly related to the acceleration parameter(s) of
 the problem.
 As is well known, such effects have been considered for the first time
 by Unruh \cite{u}, who introduced the concept of {\em quantum detector} and
 derived the vacuum temperature, $T_V =\frac{\hbar}{2\pi c k_B}a$, for a linear
 accelerating one, and by Davies \cite{dav}, who used a mirror model, in
 the mid-seventies. Hawking effect \cite{hw} was shown to be of the same type.
 They are generally considered amongst the most fundamental things
 Physics is telling us.

  Traditionally, there are two basic physical
 mechanisms related to the quantum radiative processes \cite{ddc}. One is
 directly related to the {\em randomness} of the interaction of the
 particles with  the vacuum fluctuations, the other is the so-called
 electromagnetic self interaction, better known as
  {\em radiation reaction}. Here,
  we shall consider only the {\em randomness} aspect of the vacuum noise.
The general problem to study is the coordinate dependence of the
spectrum of vacuum fluctuations as seen by quantum detectors/particles
moving along classical trajectories. This noise spectrum is a density
of states times the Fourier transform of the autocorrelation function
(Wightman two-point function) of a quantum field along the classical
trajectory. For a relativistic, linear accelerated detector the
scalar vacuum noise has an {\em exact} thermal representation. This is
 a remarkable theoretical result because one could identify the
temperature parameter directly as a scaled kinematical invariant
 (the proper acceleration); the rest of physics (quantum mechanics,
 relativity and statistical mechanics) enters as a scaling factor of
the corresponding fundamental constants. One would like to
 prove this ``beautiful truth" by experiment, and indeed
a number of proposals to detect such quantum ``thermal"
effects have been made in the past (for a review, see \cite{ro}). One of
them belongs to Rogers \cite{1} and refers to the very successful Geonium
Physics.


The idea of Rogers is to place a small superconducting
 PT containing a single electron in a microwave cavity and perform a resonant
transfer of the cyclotron
 vacuum noise at the relativistic-shifted axial frequency to a cavity mode.

Rogers' trap is somewhat unusual compared to
the common geonium traps.
The cap electrode separation is $2z_{0}$ = 2 mm, the electrode potentials
$\pm U _{0} =\pm 10$ kV, and the magnetic field is $B=150$ kG,
whereas representative values of parameters for geonium traps are
electrode separations of tens of mm, electrode potentials of several
volts, and magnetic fields of 1-2 T (Dehmelt has used a trap with
$2z_0$ = 8mm, quadrupole potential of several volts and magnetic field of 5 T).
The 15 T of the Rogers trap
are attainable with superconducting solenoid magnets.
The microwave cavity is of
length 1 cm, radius 1.36 cm, and $Q\approx 10^{4}$.

The single electron is constrained to move in a cyclotron orbit
around the trap axis by the powerful axial magnetic field.
Rogers considered the circular proper acceleration to be
$a=6\times 10^{19} g_{\oplus}$, i.e., an electron having the velocity
$\beta=0.6$. This acceleration, if considered as linear, corresponds to a
temperature of the vacuum $T_V = 2.4 K$. The velocity of the electron is
maintained constant by means of a circularly polarized wave with the frequency
equal to the cyclotron one, compensating at the same time for
the irradiated power (the synchrotron damping width is
$\Gamma _c = \frac{4e^2\omega _c^2}{3mc^2}$). The static quadrupole
electric field of the trap
creates a quadratic potential well along the trap axis in which the electron
 oscillates. The frequency of observation is
$\omega = \gamma \omega _z$ = 10.57 GHz for the device scale and working
conditions chosen by Rogers. At this frequency, the difference in energy
densities between cyclotron (circular) noise and the universal Unruh noise are
negligible. The spectral power density of the cyclotron noise at the
relativistic-shifted axial frequency is
$\partial P/\partial f= 0.47\cdot 10^{-22}$ W/Hz. This power is
resonantly transferred to the $TM_{010}$ mode of the microwave cavity
and a most sensible cryogenic GaAs field-effect transistor amplifier
 should be used in order to have an acceptable signal-to-noise ratio
 S/N = 0.3. According to Rogers, the signal may be distinguished from
 the amplifier noise in about 12 msec.

The experiment of Rogers requires the top of electronics,
cryogenic techniques, and geonium methods.


The great advantage of Rogers' proposal over that of Bell and
Leinaas \cite{bl} resides in using a single electron instead of
electron bunches. In this way, the very complicated stochasticity of the
beam dynamics at a storage ring is avoided and a better control of the
dynamics is allowed just because of the small spatial scale of the
Penning device.

The PT problem can be defined as the interaction of the electron
with an external field consisting of a uniform magnetic field
${\bf B}=B{\bf \hat {e}}_{z}$ and a
superposed electric quadrupole potential :
$$\Phi(x,y,z)= \frac{U_{0}}{2z_{0}^{2}+ r_{0}^{2}}(2z^{2}-x^{2}-y^{2})~.
\eqno(1)
$$
Details of the practical achievement of this combination of fields are
given in \cite{p}. The electrostatic field is $E=\nabla \Phi$. $U_{0}$
is the potential difference between a one-sheeted hyperboloid which is
the ring electrode and a two-sheeted hyperboloid forming the cap
electrodes, $r_{0}$ is the inner radius of the ring electrode and
$2z_{0}$ is the distance between the two end-cap electrodes.
 In the ideal case (i.e., perfect cylindrical geometry) there exist
three bounded motions in the trap: (i) an axial harmonic motion in a
 parabolic well along the direction of the
magnetic field, (ii) the cyclotron motion at a higher frequency,
 in the perpendicular (x,y) plane, and (iii) a magnetron motion
in the same plane at a much lower frequency. The axial and cyclotron
motions are usually excited with nearly resonant , radio-frequency
drives applied to the trap electrodes. For a single electron in the
 trap (obtained with an ``evaporation" technique due to Dehmelt and
 collaborators)
 the classical equations of motion are as follows
$$ \ddot{z} +\omega^{2} _{z} z=0~,
\eqno(2a)
$$
$$ \ddot{r} = \frac{1}{2} \omega^{2} _{z} r -i\omega_{c} \dot{r}~,
\eqno(2b)
$$
where $ r =x+iy$, $\omega^{2} _{z}= 4eU_{0}/m(r^{2} _{0} + 2 z^{2} _{0})$,
$\omega _{c} =eB/mc$.
The solutions are
$$z=r_{z}cos\omega_{z} t
\eqno(3a)
$$
and
$$ r = r_{c}e^{-i\omega^{'} _{c} t} + r_{m}
e^{-i\omega_{m} t}~,
\eqno(3b)
$$
where
$$\left\{ \begin{array}{ll}
        \omega^{'} _{c}\\
        \omega_{m}
        \end{array} \right\}
          =\frac{1}{2}\omega_{c} \pm
   [(\omega_{c} /2)^{2}-(\omega^{2} _{z})/2]^{1/2}~.
\eqno(4)
$$

The amplitudes $ r_{m}$, $ r_{c}$, $ r_{z}$, as well as
some possible phases are determined by the initial conditions. The
motion is stable only for $\omega ^{2} _{c} > 2\omega ^{2} _{z} $, see the
rhs of Eq.~(4),
which is the so-called trapping condition, implying periodic solutions.
A typical xy orbit for the case $\omega _{m} \ll \omega ^{'} _{c}$,
$ r_{m}  < r_{c}$ is shown in Fig.~1 of Ref.~\cite{x}.
Only under such conditions, the magnetron motion can be interpreted as an
$\vec{E}\times \vec{B}$ drift of the center of the cyclotron orbit
 around the trap axis.

The electron motion in the trap can be analyzed by many procedures in
analytical mechanics \cite{k}. However, we are interested in the transition
rates of a quantum electron due to the vacuum fluctuations affecting
its classical trajectory. Unfortunately, the classical trajectory in
the trap is not a simple one. From the point of view of the circular
(cyclotron) vacuum noise we are not in the ideal situation;
the magnetron drift is also there, though usually considered as negligible,
or diminished by pumping at $\omega _z+\omega _m$ \cite{73}.
Besides, one has to take care of all
the other, more common sources of noise (for a discussion of the axial
Brownian noise see Ref.~1b).
Even in the case of pure cyclotron motion, it
is well known that the circular vacuum noise is not at all
universal thermal ambience \cite{kor}. Rogers has used the
formula for the spectral energy density of a massless scalar field as
given by Kim, Soh and Yee (KSY) \cite{kor}, i.e.,
$$
\Bigg[\frac{de}{d\omega}\Bigg]_{\omega=\gamma \omega _z}=
\frac{\omega ^3}{\pi ^2}\left(\frac{1}{2} +
\frac{1}{2\gamma ^2 r}\sum_{n=0}^{\infty}v^{2n}f_n(r)\right)~,
\eqno(5)
$$
where $r=\frac{\omega}{\gamma \omega _0}$, $v$ is the electron velocity,
$\gamma$ is the Lorentz factor, and $f_n$ is a sum over an index $k$
of step functions of argument $n-k-r$, indicating the excitation of zero-point
field modes at frequencies of the type $p\omega _c \pm \omega _z$, with
$p$ a positive integer; in
other words, the vacuum noise at $\gamma \omega _z$ is a sum of cyclotron
axial-displaced harmonics.

 The KSY parametrization of the circular spectrum gives a quasi-continuous
spectrum which is similar to the black-body one only at low frequencies and
for the first few terms in the sum (Rogers considered the first four terms).
The real PT
cyclotron-axial-magnetron motion produces a more complicated vacuum noise,
that one may call the Penning vacuum noise (PVN). Moreover, Levin, Peleg, and
Peres \cite{lpp} have shown that further Casimir corrections might be
important. As a matter of fact, cavity effects have been thoroughly
investigated in the literature in view of the decade-old debate on their
direct substantial shifting of the spin magnetic moment \cite{lsb}.
Also, according to Becker, Wodkiewicz and Zubairy \cite{bwz},
a tiny squeezing effect of the cavity modes is possible just because of the
presence of the electron in the cavity.

I would like now to emphasize that recently constructed
cylindrical PTs \cite{cyl}, for which the
trap itself is a microwave cavity, should be considered as an
important simplification of the experimental setup for detecting the
Penning Unruh-like noise. In this case, small slits ($\approx$ 0.015 cm) at
$\pm z_0$ above and below the trap center,
incorporating choke
flanges, divide the oxygen-free high-conductivity copper cavity walls
into the required Penning electrode configuration, i.e., the two end-cap
electrodes at the level
of the slits and the ring electrode (of radius $r_0$).
Furthermore, there are two
essential compensation electrodes of height $\Delta z_c/z_0$ = 0.20 placed
near the end-cap ones, on which the applied potential is tuned in
order to make the axial motion of the electron as harmonic as possible. It has
been shown \cite{cyl} that an ``orthogonalized" geometry of the trap, i.e.,
$r _0 = z_0$, reduces a lot the nonlinear axial frequency shifts.
The three motions of the hyperbolic trap, the cyclotron, axial, and magnetron
ones take place at $\approx$ 166 GHz, 63 MHz, and 12 kHz,
respectively \cite{cyl}. Moreover,
the driven axial resonance for this configuration has been
observed with almost the same signal-to-noise ratio as in hyperbolic
PTs. By means of these cylindrical cavity-traps, a direct coupling to
the cavity modes may be achieved, especially in the common weak coupling
regime, where the cyclotron oscillator and the cavity mode cannot form
normal modes, and thus supplementary nonlinear effects are not coming
into play. The cylindrical Penning Unruh-like noise will be a tiny form of
radiative cooling of the electron oscillator.
The cylindrical $TM_{010}$ mode is essentially a
Bessel function of zero order in the radial direction without nodes along the
$z$-axis. The frequency of $TM_{010}$ is given by $\omega _{010}=
\frac{c\xi _{01}}{\rho _0}$, where $\xi _{01}$= 2.405 is the first zero of
the Bessel function of zero order.
The price to pay in the case of the cylindrical trap is a loss
in the quality of the electrostatic quadrupole potential, despite the
compensating electrodes helping partial control of this problem.

I now pass to some comments on other related problems in Penning trap physics.
Some time ago Fern\'andez and Nieto \cite{fn} used the Weyl-Wigner-Moyal
phase space formalism to calculate the energy levels of a spinning
charged particle in a PT. Here we wish to point out that the study of
the PT phase space is a problem one should place in the more
general context of Chaos Science \cite{ch}.
In phase space the uncertainty principle replaces the
continuum of classical states within a volume $\hbar ^{N}$
 (where N is the number of degrees of freedom of the system) by a
single quantum state, and
as such, quantum chaos is less `powerful' than the classical one. In a
 certain sense, to go beyond quantum mechanics, at least in phase space,
 one might think of more accurate topologies defining the proximity of
 points (states) \cite{l}. In this way, one will have an intermediate
 picture between quantum chaos and the standard geometric (classical)
 chaos. The geometric chaos is a mathematical one devoided of physical
 reality. For the physical world it is the ideal-limiting case never
 satisfied by real measurements. However, one would like to estimate as
 best as possible how far we are from the ideal case in an experiment.
 In this sense, the proximity topology might help progress in a better
 evaluation of experimental precision.

One should also keep in mind that in the case of a particle
moving in a plane to which a magnetic field is applied in the
normal direction, the momentum operators cease to commute:
$$[p_{1},p_{2}]=ieB\hbar~.
\eqno(6)
$$
This introduces a cellular structure in the momentum plane, which
becomes divided into Landau cells of area proportional to $eB\hbar$.
Thus, in the PT case, the Landau cells are to be found in the momentum plane
normal to the axial magnetic field.

We also recall that in the semiclassical limit the magnetic field
determines not only the classical trajectory of the particle through
the Lorentz force (dynamical effect), but also contributes to the
phase accumulated along the trajectory through the line integral
of the vector potential along it (geometric effect) \cite{fb}.


In conclusion, as far as the electromagnetic trapping is considered as
one of the most
precise tools at our disposal, one might think that extremely tiny
fundamental effects are best accessible by means of such experimental
techniques. However, one should be fully aware of the host of more standard
effects which are also involved in the trap laboratory.
In my short inspection of Rogers' proposal I put forth a
number of suggestions that presumably are worthy to be pursued in future
more detailed studies. The use of cylindrical traps may contribute to
turn not only Rogers' proposal into a more realistic one, but also other
even more exotic proposals \cite{kar}.

Finally, if one takes the viewpoint that the circular Unruh noise is
actually the common electromagnetic radiation of an accelerated electron
(this is the standpoint of the present
author too) then the discussions of the vacuum noise are still of some worth,
this time aimed merely at introducing a new temperature parameter in the
PT physics.

\section*{ Acknowledgments}

This work was partially supported by the CONACyT Project 4868-E9406.


\end{document}